\documentclass[conference]{IEEEtran}
\IEEEoverridecommandlockouts
\usepackage{cite}
\usepackage{amsmath,amssymb,amsfonts}
\usepackage{algorithmic}
\usepackage{graphicx}
\usepackage{textcomp}
\usepackage{xcolor}
\def\BibTeX{{\rm B\kern-.05em{\sc i\kern-.025em b}\kern-.08em
    T\kern-.1667em\lower.7ex\hbox{E}\kern-.125emX}}
\begin{document}

\title{Analytical Inverter-Based Distributed Generator Model for Power Flow Analysis}

\author{\IEEEauthorblockN{Naeem Turner-Bandele$^{1}$, Amritanshu Pandey$^{1,2}$, and Larry Pileggi$^{1}$}
\IEEEauthorblockA{\textit{1. Electrical and Computer Engineering Dept. 2. Engineering and Public Policy Dept.} \\
\textit{Carnegie Mellon University}\\}}

\maketitle

\begin{abstract}
Quantifying the impact of inverter-based distributed generation (DG) sources on power-flow distribution system cases is arduous. Existing distribution system tools predominately model distributed generation sources as either negative PQ loads or as a PV generator and then employed a PV-PQ switching algorithm to mimic Volt/VAR support. These models neglect the unique characteristics of inverter-based distributed generation sources, have scalability and convergence issues, and are ill-suited for increasing solar penetration scenarios. This work proposes an inverter-based DG model accounting for the inverter's topology, sensing position, and control strategies. The model extends recently introduced analytical positive sequence generator models for three-phase studies. The use of circuit-simulation based heuristics help achieve robust convergence. Simulation of the PG\&E prototypical feeders using a prototype solver demonstrate the model's accuracy and efficacy.
\end{abstract}


\section{Introduction}
Declining costs and increasing environmental concerns fuel rising penetration rates of distributed solar energy on the U.S. electric distribution system. Solar photovoltaics accounted for 37\% of all new electricity generating capacity additions in the first half of 2020 \cite{Perea2020}. Projections for the future illustrate a trend towards more worldwide solar generation continuing over the next several decades despite the short-term declines caused by the COVID-19 pandemic \cite{Perea2020}. 

Historically, rising penetration prompted concerns over solar generation's impact on voltage quality, power losses, and other voltage regulation devices. Lately, those concerns have diminished. Distributed solar energy is interfaced with the electric grid using modern inverters that employ much more sophisticated control strategies. These inverters can provide system support under a variety of conditions. Despite these improvements, new challenges surface related to the control, protection, and interaction of distributed solar with other energy sources. To tackle these challenges, modeling and analysis of inverter-based distributed generation must improve.

In a three-phase power flow analysis for unbalanced distribution grids, solar DGs are generally modeled either as negative PQ loads or PV generators. A PV-PQ switching algorithm simulates an inverter's ability to provide regulation-mandated reactive power support and ensures that the generator's physical limits are incorporated. These models neglect the unique characteristics of inverter-based distributed generation sources \cite{NorthAmericanElectricReliabilityCorporation2017, STREZOSKI2019, Abdel-GalilAhmedEBAbu-Elanien2007, Kamh2011} and have scalability and convergence issues \cite{Massignan2017}. Worse, they are ill-suited for increasing solar penetration scenarios \cite{NorthAmericanElectricReliabilityCorporation2017, STREZOSKI2019}. The North American Electric Reliability Corporation (NERC) advises that DG modeling should consider their dynamic characteristics and steady-state output \cite{NorthAmericanElectricReliabilityCorporation2017}. 

Recent work reveals steady-state models better accounting for the properties and characteristics of inverter-based distributed energy resources are possible \cite{Hwang2016, Kamh2011, Radatz2020}. In \cite{Radatz2020}, the authors present the OpenDSS model, which accounts for photovoltaic array attributes, inverter properties, and inverter controls but does not incorporate specific inverter control strategies and implements control functions via a control loop rather than implicitly. In \cite{Hwang2016}, the authors propose models that consider the topology, sensing position,  output filter, and control strategy of the inverter-based distributed generators (IBDG). This work successfully demonstrated convergence for small cases with low IBDG penetration. However, this work neglected inverter current limits, Volt/VAR control, and failed to demonstrate realistic-size networks' scalability and convergence.

Conducting a steady-state analysis of an IBDG in the distribution system requires a three-phase power flow or load flow analysis. Several methods exist to solve the nonlinear equations, including the forward-backward sweep (FBS) \cite{Srinivas2000}, Gauss-Seidel (GS) \cite{Teng2002}, or three-phase current-injection (TCIM) \cite{Garcia2000} techniques. Each method attempts to solve a system of nonlinear equations until convergence to a solution. Despite their historical prominence, all of these methods suffer from either divergence, convergence to incorrect solutions, or difficulty at scale [12]. Recently, an equivalent circuit approach shows improved convergence by using circuit simulation-based heuristics \cite{Pandey2018}. The equivalent circuit approach maps power system components to equivalent circuits, using current and voltage as the state variables, and then uses circuit simulation heuristics and homotopy methods to achieve robust convergence  \cite{Pandey2018}.

This work proposes an inverter-based DG model that builds upon prior modeling efforts. It introduces overcurrent limits and implements Volt/VAR control that adheres to current regulations. To implement the new additions, the model uses equivalent circuit models to construct an analytic generator model that is first-order continuous \cite{Agarwal2018}. The improved IBDG model achieves robust convergence in the corresponding three-phase power flow simulation. Results indicate that if using a high percentage of inverter-based devices on large systems, standard models might misrepresent the actual grid state, potentially impacting decision-making. 

The organization of the remainder of this paper is as follows. Section II discusses the equivalent circuit formulation, recent efforts on improved modeling of IBDGs, and Volt/VAR control. Section III extends prior work by incorporating overcurrent control and implicit Volt/VAR control. The results are in Section IV. Conclusions and future directions are in Section V.

\section{Background}

\subsection{Three-Phase Power Flow Equivalent Circuit Formulation}
Recently, the authors of \cite{Pandey2018} introduced an equivalent circuit approach for power flow and three-phase power flow problems. Within this approach, all power systems components are mapped to an equivalent steady-state circuit using current and voltage as the primary state variables. Solving the corresponding nonlinear system equations requires using the Newton-Raphson method on all three phases. An example of this method for a three-phase PV bus or generator is as follows.

A wye-connected three-phase generator can be modeled as three complex current sources connected by a neutral. On their own, the complex current sources and their corresponding equations are insufficient for applying NR since the complex conjugate function that governs each current source is non-differentiable. Applications of Euler’s  formula and circuit theory show that we can split a complex, steady-state circuit in terms of its real and imaginary components. Thus, all PV complex current sources can be divided into a real ($I_{\phi, RG}$) and imaginary current source ($I_{\phi, IG}$) to create analytic functions. The corresponding equations of the current sources are:

\begin{equation}
I_{\phi, RG} = \frac{P_{\phi,G}V_{\phi, RG} + Q_{\phi,G}V_{\phi, IG}}{(V_{\phi, RG})^2  + (V_{\phi, IG})^2}
\end{equation}
\begin{equation}
I_{\phi, IG} = \frac{P_{\phi,G}V_{\phi, IG} - Q_{\phi,G}V_{\phi, RG}}{(V_{\phi, RG})^2  + (V_{\phi, IG})^2}
\end{equation}
Here $\phi$ indicates the phase A, B, or C of the power system.

\begin{figure}[htbp]
    \centering
    {\includegraphics[width=0.9\columnwidth]{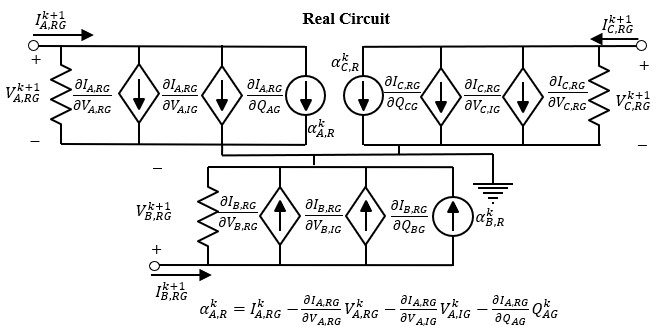}} 
    \caption{Real circuit for a three-phase generator. $\alpha^k_{B,R}$, and $\alpha^k_{C,R}$ are not shown.}
    \label{fig:RealPVCkt}
\end{figure}

We can linearize these equations via a first-order Taylor expansion to represent the NR algorithm's primary step. The linearized real circuit for the three-phase generator is in Figure \ref{fig:RealPVCkt}. A similar circuit, not shown here, is also constructed for the imaginary portion.

A standard three-phase PV bus also requires three additional equations to handle the unknown, $Q_{\phi, G}$, and to ensure that generator voltage magnitudes maintain their setpoints $(V_{\phi, G})$. 
\begin{equation}
        \label{eq:volt_control}
V_{\phi,G} = (V_{\phi, RG})^2 + (V_{\phi, IG})^2
\end{equation}
The voltage control equations are linearized just as the generator current source equations were.

The benefits of this formulation are that it allows power systems to take advantage of circuit simulation heuristics and homotopy methods that are known to improve and guarantee convergence \cite{Pandey2018}. The circuit formulation and representation are vital in achieving robust convergence for large-scale cases, ill-conditioned systems, and high distributed generation areas. 

\subsection{Inverter-Based Distributed Generator Modeling}

Traditionally, inverter-based generators are modeled abstractly as fixed PQ buses or as voltage-controlled PV buses. These representations are misleading, given that IBDGs do not function like conventional generators. Further, these models assume balanced operating conditions, which are not the realities of the distribution system. As shown in \cite{Hwang2016},\cite{Teodorescu2011}, the phase outputs of a DG depend on both the positive and negative sequence components under unbalanced conditions.

Steady-state models that consider the operational characteristics and control of a three-wire current-controlled voltage source inverter (CCVSI) distributed generator are in \cite{Hwang2016}. In this formulation, the authors represent the CCSVI as a three-phase current source connected to some impedance whose output current is determined by the inverter's sensing position and control strategy. Using instantaneous power theory and circuit analysis, the authors derived 16 DG combinations based on the positive-negative sequence control (PNSC), average active-reactive control (AARC), and balanced positive sequence control (BPSC) strategies.

The governing complex and nonlinear IBDG current equation derived in \cite{Hwang2016} is: 
\begin{equation}
\label{eq:4}
    I_{\phi, G} = I_{Ref}\alpha - V_{\phi, G}\beta
\end{equation}
where $\alpha$ and $\beta$ are impedances specified by the inverter's sensing position. 

The reference current, $I_{Ref}$ varies depending on the control strategy used. While \cite{Hwang2016} uses the PNSC, BPSC, and IARC strategies, we use the FPNSC strategy. The FPNSC equation is \cite{Teodorescu2011}:
  \begin{equation}
  \label{eq:5}
    \begin{split}
            I_{Ref} &= P_{3G}\bigg(\frac{ k_{1} V^{+}_{\phi, G}}{|V^{+}_{\phi, G}|^2} + \frac{(1 - k_{1}) V^{-}_{\phi, G}}{|V^{-}_{\phi\perp, G}|^2} \bigg)\\ &+ Q_{3G}\bigg(\frac{ k_{2} V^{+}_{\phi\perp, G}}{|V^{+}_{\phi, G}|^2} + \frac{ (1 - k_{2}) V^{-}_{\phi\perp, G}}{|V^{-}_{\phi, G}|^2}\bigg)
    \end{split}
   \end{equation}
Here $P_{3G}$ is the total three-phase active power, $Q_{3G}$ is the total three-phase reactive power, $k_1$ and $k_2$ are flexible weights used to regulate sequence component's contribution, and $\perp$ denotes orthogonal voltages. Note, $V_\perp = -jV$.

The models in \cite{Hwang2016} converged for small test cases using FBS; however, the authors did not demonstrate robust convergence for large-scale cases and cases with high distributed generation penetration. Additionally, the models in \cite{Hwang2016} do not incorporate current-limits and voltage-reactive control of a DG. A comprehensive distributed generation model must also consider the control modes of a DG \cite{Teodorescu2011}.

\subsection{Voltage-Reactive Power Control}
Reactive power support and control is crucial to any IBDG model. Per IEEE 1547, IBDGs must be able to supply reactive power support when necessary \cite{IEEEStandardAssociation2018}.  Discontinuous approaches, such as PV/PQ switching, have long been employed to simulate reactive power control. However, this method has significant problems with convergence \cite{Sarmiento2018, Agarwal2018}. Further, it does not accurately represent Volt/VAR control's behavior in an IBDG since the Volt/VAR control of an inverter is more piecewise in nature \cite{IEEEStandardAssociation2018, Sarmiento2018}. Figure \ref{fig:AVR_Droop} illustrates the difference between the two approaches. 

\subsubsection{PV/PQ Switching}
\begin{figure}[!ht]
    \center
   {\includegraphics[width=0.6\columnwidth]{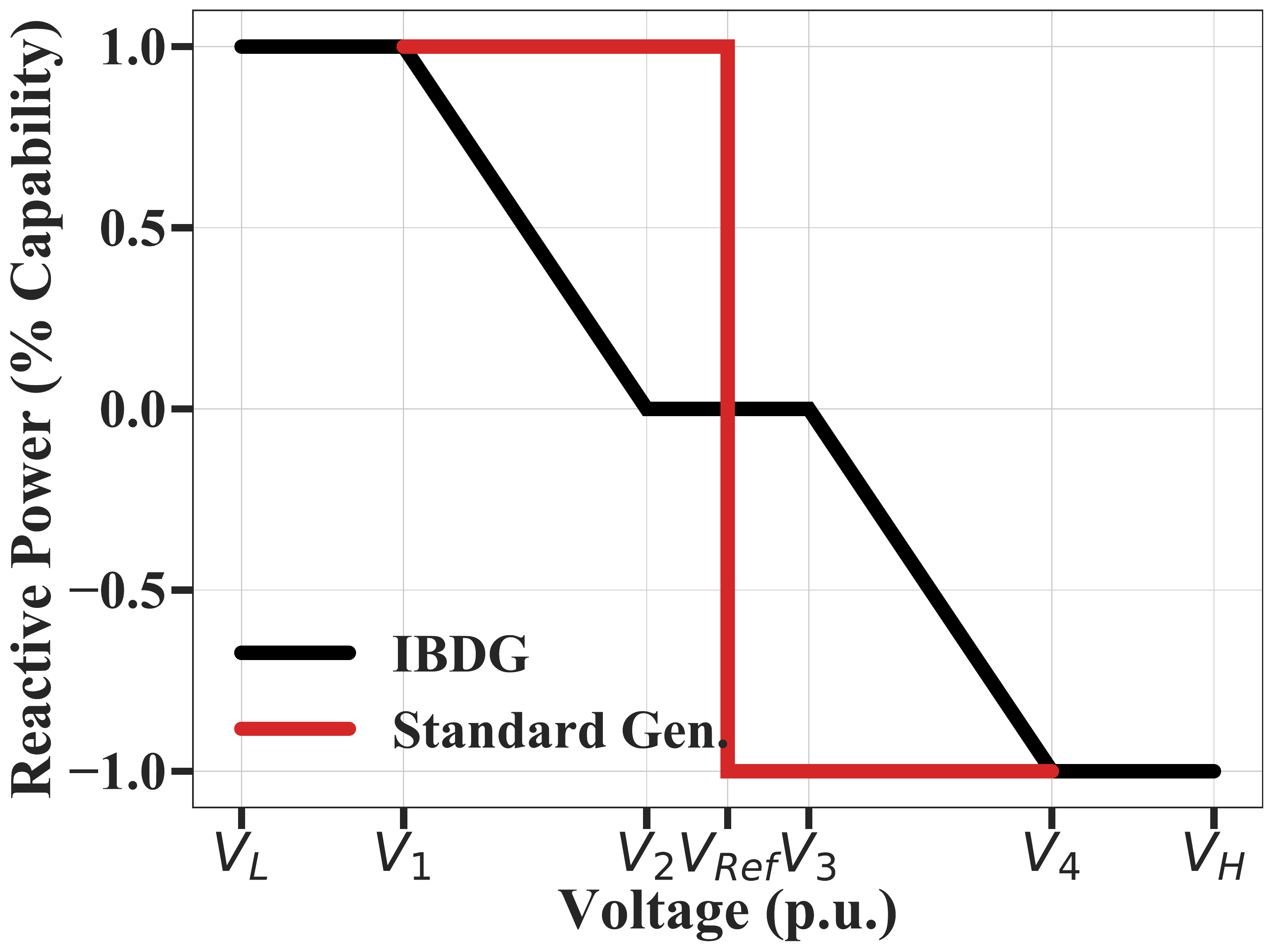}}
    \caption{Standard generator Q-V curve compared against that of an inverter-based generator's response.}
    \label{fig:AVR_Droop}
\end{figure}

PV/PQ switching involves using an outer iteration loop during power flow simulation to alternate a generator between a PV bus model and a PQ bus model. This alternating means that the reactive power Q goes from an unknown to a fixed quantity across these outer loop iterations. The objective is to maintain the generator within its reactive power limits while staying along the region identified by the red curve in Figure \ref{fig:AVR_Droop}.

\subsubsection{Piecewise Models}
Instead of the discontinuous method, a piecewise model more accurately mimics the  Volt/VAR control behavior in a distributed generator. Traditionally, piecewise models have shown difficulties with convergence due to discontinuities at function bounds. Numerous attempts have been made at developing smooth, continuous piecewise functions for implicit use in power flow models, including \cite{Allison2019, Agarwal2018}.

Conversely, in \cite{Agarwal2018}, the authors demonstrate implicit modeling of generator limits by representing the limits as a segmented piecewise continuous function. Incorporating small quadratic regions (or patches) in this model ensures the function remains continuously differentiable, and first-order derivatives match at intersecting points. A first-order continuous function enables the use of NR without the convergence issues of piecewise discontinuous models. Using relaxation and homotopy methods, the authors achieved robust convergence for a variety of large-scale test cases.

\section{Three-Phase Steady State IBDG Model}

\subsection{Equivalent Circuit Representation}
By splitting the complex current of \eqref{eq:4} in terms of its positive and negative sequence components and then using the equivalent circuit formulation procedures, we can obtain the real and imaginary currents for an IBDG using the FPNSC control strategy. Below is the real current for the positive sequence. In \eqref{eq:FPNSCR} $\alpha_R$, $\alpha_I$, $\beta_R$, and $\beta_I$ are the real and imaginary sensing position impedances. An imaginary positive sequence current and real and imaginary currents for the negative sequence also exist but are not shown here.

 \begin{equation}
        \label{eq:FPNSCR}
        \begin{split}
            I^{+}_{\phi, RG} &= k_{1} P_{3G} \bigg(\frac{\alpha_R V^{+}_{\phi, RG}  - \alpha_{I} V^{+}_{\phi, IG}}{(V^{+}_{\phi, RG})^2 + (V^{+}_{\phi, IG})^2}\bigg) \\
            &+ k_{2} Q_{3G}\bigg(\frac{\alpha_I V^{+}_{\phi, RG}  + \alpha_{R} V^{+}_{\phi, IG}}{(V^{+}_{\phi, RG})^2 + (V^{+}_{\phi, IG})^2}\bigg) \\
            &+ \beta_{R}V^{+}_{\phi, RG} - \beta_{I}V^{+}_{\phi, IG}
        \end{split}
    \end{equation}
A first-order Taylor expansion of \eqref{eq:FPNSCR} is taken to linearize the function and express the equation in terms of its equivalent circuit components following Section II's procedures. The linearized real circuit representing \eqref{eq:FPNSCR} is like Figure \ref{fig:RealPVCkt}. Not shown here, three additional circuits also exist to represent the imaginary positive sequence and the inverter's real and imaginary negative sequence.

\subsection{Current Limitation and Reactive Power Limits}
The phase-to-phase current outputs of the IBDG are not guaranteed to be balanced. In a three-phase unbalanced system, overcurrent in any system phase can trigger disconnections from the point of common coupling \cite{Teodorescu2011}. We can ensure that a disconnect does not occur by determining the permitted instantaneous current limits. The maximum current must be calculated for each control strategy since no single expression exists \cite{Teodorescu2011}. 

For FPNSC, the derivation of the maximum current and the subsequent reactive power limits is as follows. First, we transform the FPNSC reference current to the stationary reference frame. Then, using the stationary representation, we can find the peak current for phase A, B, and C in terms of a singular equation using the methods in \cite{Teodorescu2011}.
\begin{align}
        \label{eq:IPK}
    \begin{split}
        I^2_{pk} &= (P_{3G} C_1 \cos{\gamma} - Q_{3G} C_2 \sin{\gamma})^2 \\
        &+ (-Q_{3G} C_3 \cos{\gamma} - P_{3G} C_4 \sin{\gamma})^2
    \end{split}
\end{align}
where $\gamma$ is the rotated positive and negative sequence voltage angle difference per phase, $C_1 = \frac{k_{1}}{|V^{+}_{\phi,G}|} + \frac{(1 - k_{1})}{|V^{-}_{\phi,G}|}$, $C_2 = \frac{k_{1}}{|V^{+}_{\phi,G}|} - \frac{(1 - k_{1})}{|V^{-}_{\phi,G}|}$, $C_3 = \frac{k_{2}}{|V^{+}_{\phi,G}|} + \frac{(1 - k_{2})}{|V^{-}_{\phi,G}|}$, and $C_4 = \frac{k_{2}}{|V^{+}_{\phi,G}|} - \frac{(1 - k_{2})}{|V^{-}_{\phi,G}|}$. Using this expression for the maximum current, we can estimate the reactive power limits and ensure that the IBDG currents do not exceed their limits.

We can use the maximum current to obtain the reactive power expression considering simultaneous active and reactive power delivery or the maximum reactive power. Given that active power is constant for this representation of an inverter, we only present the latter.


After expanding \eqref{eq:IPK}, setting $P_{3G}$ equal to zero, and manipulating the resulting algebraic equation, we obtain a final expression for the maximum reactive power \cite{Teodorescu2011}:


\begin{equation}
    Q_{3G, max} = \sqrt{\frac{I^2_{pk}|V^{+}_{\phi,G}|^2 |V^{-}_{\phi,G}|^2}{B}}
\end{equation}
\begin{align}
    \begin{split}
            B &= k^{2}_{2}|V^{-}_{\phi,G}|^2 + (1-k_{2})^2|V^{+}_{\phi,G}|^2 \\
            &- 2k_{2}(1 - k{2})|V^{-}_{\phi,G}||V^{+}_{\phi,G}|\cos{2\gamma}
    \end{split}
\end{align}
Since $\gamma$ has three different possible values, the above expression will produce three different values for $Q_{3G, max}$. Within the IBDG model, the lowest possible $Q_{3G, max}$ value is used, and its corresponding phase is taken into account. Over this minimum $Q_{3G}$, the current limits of all phases will be exceeded \cite{Teodorescu2011}.

\subsection{Volt/VAR Control}
\begin{figure}[!ht]
    \centering
    {\includegraphics[width=0.6\columnwidth]{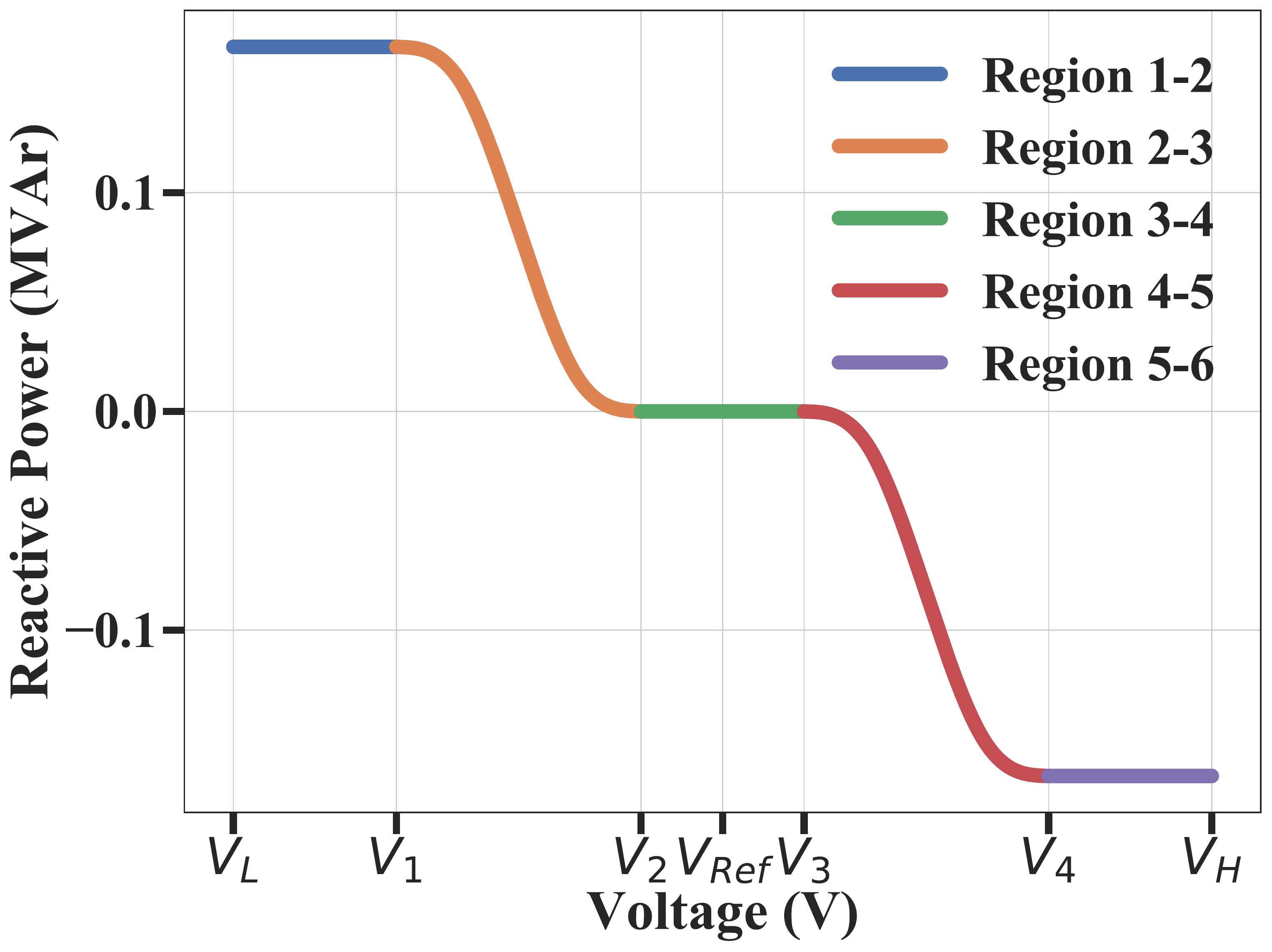}}
    \caption{Example piecewise Volt/VAR curve for an IBDG.}
    \label{fig:piecewise}
\end{figure}

The IBDG model includes a Volt/VAR control that better aligns with reactive control in inverters. A smooth and continuous piecewise patching function expressing the reactive power in terms of voltage was created. The Volt/VAR function's curved regions were achieved using a natural cubic spline (a composite polynomial function). The splines serve as transitions between the linear voltage regions. A natural cubic spline is one where the function is twice continuously differentiable and ensures that the derivatives at the endpoints are zero. An example of the function is shown graphically in Figure \ref{fig:piecewise}. Using homotopy methods from \cite{Agarwal2018} ensures robust convergence.

\section{Results}
We incorporated the IBDG model into the three-phase circuit-based power flow solver in \cite{Pandey2018} that is called SUGAR. Testing of the IBDG occurred in two stages. First, we verified the model’s accuracy by comparing the steady-state power flow results with synthetic measurements generated by electromagnetic transient simulation in EMTP-RV for a single-node test case. We then assessed how existing models’ inaccuracies could affect future electric distribution systems with significant photovoltaic penetration by studying the PG\&E prototypical feeders.

\subsection{Single-Node Validation}
\begin{figure}[!ht]
    \centering
    {\includegraphics[width=0.6\columnwidth]{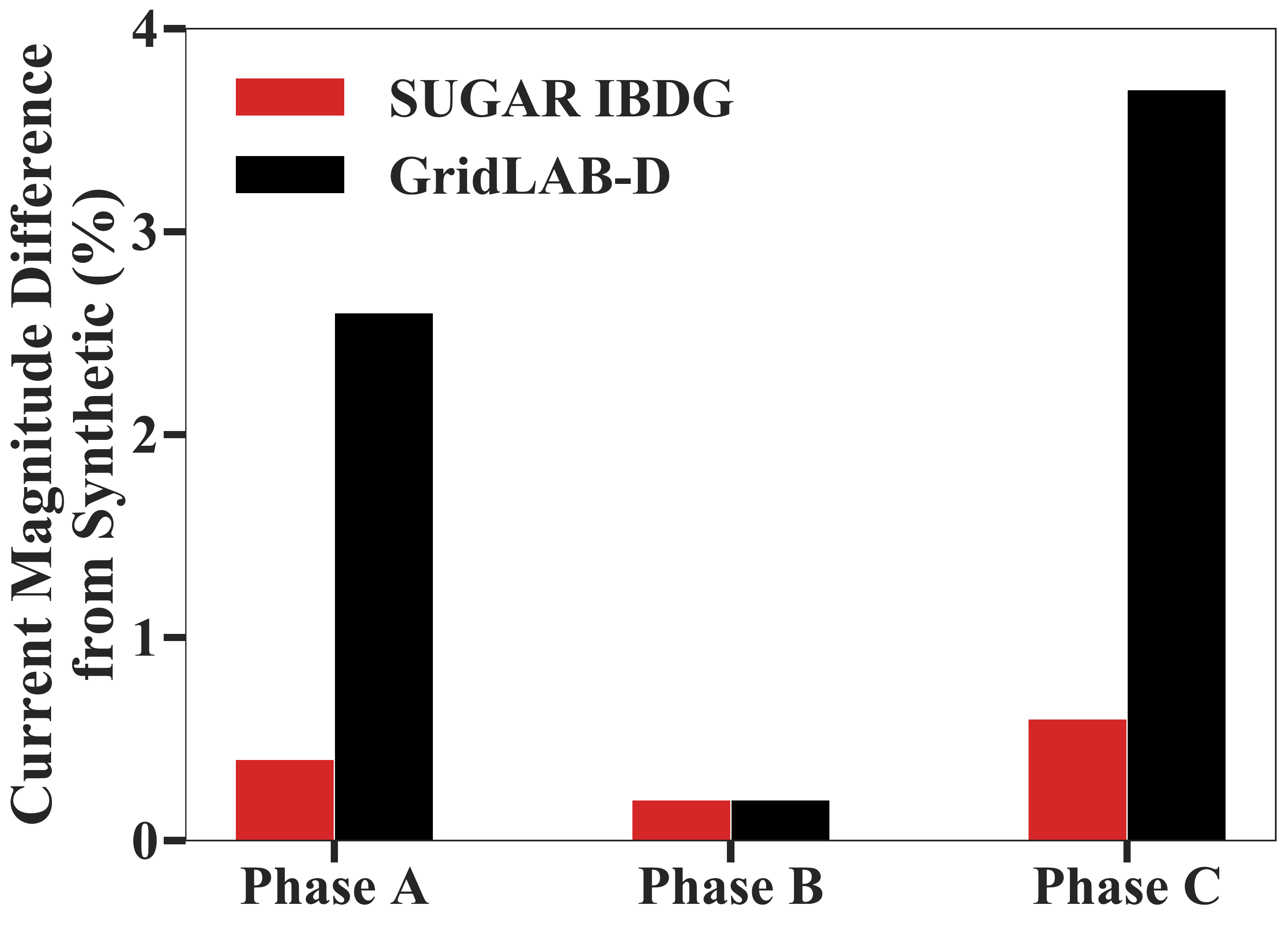}}
    \caption{The current magnitude percent difference of the IBDG and GridLAB-D inverter models from the synthetic measurements generated in EMTP-RV.}
    \label{fig:current-diff}
\end{figure}
As a baseline, we first examined how the IBDG model performed against the GridLAB-D inverter and a synthetic measurement generated in EMPT-RV for a single-node test case. In the single-node test case, the solar distributed generator connects directly to a slack node. The EMTP-RV transient simulation included a fully dynamic solar panel model, the inverter, and an FPNSC control system. The transient analysis is run from $t=0$ to an approximate steady-state at $t=5$ seconds.  

\begin{table}[ht]
\centering
\caption{Overview of the simulated PG\&E prototypical feeders.}
\resizebox{0.55\textwidth}{!}{%
\begin{tabular}{|l|r|r|r|ll}
\cline{1-4}
\textbf{Feeder   Name} & \multicolumn{1}{l|}{\textbf{BU0001}} & \multicolumn{1}{l|}{\textbf{HL0004}} & \multicolumn{1}{l|}{\textbf{TMP009}} & \textbf{} & \textbf{} \\ \cline{1-4}
\textbf{\# Buses}         & 160      & 2,091   & 5,195       & \multicolumn{1}{r}{} & \multicolumn{1}{r}{} \\ \cline{1-4}
\textbf{Load (MVA)}       & 1.0     & 41.7   & 1.4        & \multicolumn{1}{r}{} & \multicolumn{1}{r}{} \\ \cline{1-4}
\textbf{Loading Scenario} & Off-Peak & On-Peak & Nominal     &                      &                      \\ \cline{1-4}
\textbf{PV Location}      & End      & Center  & Distributed &                      &                      \\ \cline{1-4}
\end{tabular}%
}
\label{tab:feeder-overview}
\end{table}
Figure \ref{fig:current-diff} compares the current magnitude difference between the IBDG, GridLAB-D inverter, and the EMTP-RV steady-state solution. The results illustrate that the IBDG is closer to the synthetic measurement, with errors under a percent. Conversely, the error between GridLAB-D and the synthetic measurement is less than a percent only on phase B. In a small system, this error may be manageable. However, on a larger grid with significant IBDG penetration, the IBDG error will increase overall system error. A large system error could mean that steady-state analysis tools report grid instability when the grid is stable and might affect operations and planning decisions.

\begin{table*}[]
\centering
\caption{Voltage profile of the PG\&E prototypical feeders in GridLAB-D and SUGAR}
\label{tab:feeder-results}
\resizebox{0.8\textwidth}{!}{%
\begin{tabular}{lllllllllll}
\cline{2-11}
\multicolumn{1}{l|}{\textbf{}} &
  \multicolumn{5}{c|}{\textbf{SUGAR IBDG}} &
  \multicolumn{5}{c|}{\textbf{GridLAB-D}} \\ \cline{2-11} 
\multicolumn{1}{l|}{\textbf{}} &
  \multicolumn{1}{l|}{$V_{min}$} &
  \multicolumn{1}{l|}{$V_{max}$} &
  \multicolumn{1}{l|}{$V_{mean}$} &
  \multicolumn{1}{c|}{$V_{diff.}$} &
  \multicolumn{1}{l|}{$V_{unb.}$} &
  \multicolumn{1}{l|}{$V_{min}$} &
  \multicolumn{1}{l|}{$V_{max}$} &
  \multicolumn{1}{l|}{$V_{mean}$} &
  \multicolumn{1}{c|}{$V_{diff.}$} &
  \multicolumn{1}{l|}{$V_{unb.}$} \\ \hline
\multicolumn{1}{|l|}{\textbf{BU0001}} &
  \multicolumn{1}{l|}{0.96} &
  \multicolumn{1}{l|}{1.00} &
  \multicolumn{1}{l|}{0.96} &
  \multicolumn{1}{l|}{0.03} &
  \multicolumn{1}{l|}{0.3\%} &
  \multicolumn{1}{l|}{1.00} &
  \multicolumn{1}{l|}{1.29} &
  \multicolumn{1}{l|}{1.08} &
  \multicolumn{1}{l|}{0.22} &
  \multicolumn{1}{l|}{1.5\%} \\ \hline
\multicolumn{1}{|l|}{\textbf{HL0004}} &
  \multicolumn{1}{l|}{0.79} &
  \multicolumn{1}{l|}{1.00} &
  \multicolumn{1}{l|}{0.90} &
  \multicolumn{1}{l|}{0.09} &
  \multicolumn{1}{l|}{3.4\%} &
  \multicolumn{1}{l|}{0.88} &
  \multicolumn{1}{l|}{1.29} &
  \multicolumn{1}{l|}{1.05} &
  \multicolumn{1}{l|}{0.05} &
  \multicolumn{1}{l|}{3.5\%} \\ \hline
\multicolumn{1}{|l|}{\textbf{TMP0009}} &
  \multicolumn{1}{l|}{0.98} &
  \multicolumn{1}{l|}{1.00} &
  \multicolumn{1}{l|}{0.99} &
  \multicolumn{1}{l|}{0.01} &
  \multicolumn{1}{l|}{0.5\%} &
  \multicolumn{1}{l|}{1.00} &
  \multicolumn{1}{l|}{1.29} &
  \multicolumn{1}{l|}{1.09} &
  \multicolumn{1}{l|}{0.07} &
  \multicolumn{1}{l|}{2.9\%} \\ \hline
\end{tabular}%
}
\end{table*}

\subsection{PG\&E Prototypical Feeder Evaluation}
After validating the IBDG for the case of a single node, we examined how the IBDG might perform in California's future electric distribution systems. We chose California because of the state's high solar energy potential: the Solar Energy Industries Association projects, California will install an additional 15.2 MW of solar over the next five years \cite{SolarEnergyIndustriesAssociation}. Three different types of Pacific Gas and Electric Company (PG\&E) primary distribution feeders were selected to study the IBDG model's impact. The feeders are from a  California Energy Commission and PG\&E project that studied potential voltage impacts on electric distribution circuits as photovoltaic penetration increased \cite{GridLAB-D}.

General information regarding the selected feeders is in Table \ref{tab:feeder-overview}. Each feeder was evaluated with solar generation located centrally in the feeder, at the feeder’s end, or distributed. Each feeder simulation uses either nominal, on-peak, or off-peak loading. The IBDGs are supply-driven in all studies, and the IBDG Volt/VAR curves are determined programmatically. GridLAB-D inverters use their default conditions and settings.

Table \ref{tab:feeder-results} reports the resulting comparison of IBDG and GridLAB-D inverter performance on the PG\&E feeders. In the table, $V_{min}$ and $V_{max}$ are the maximum and minimum per-unit feeder voltages, $V_{mean}$ is the mean feeder voltage, and $V_{unb.}$ is the average percent voltage unbalance in the feeder. Additionally, $V_{diff}$ represents how far the voltages at the inverter locations deviate from their reported per unit nominal voltage. For the BU0001 and TMP0009 feeders, the voltage magnitudes range was closer to the nominal using a solution with an analytic model. Feeder HL0004 shows a solution without an analytic model results in overvoltage at the feeder buses under peak loading. Conversely, using an analytic model on the HL0004 case shows the system is more likely to experience undervoltage at peak loading.   

The resulting maximum and minimum voltages on all three feeders show the advantage of incorporating a Volt/VAR control that considers the inverter's limit. The difference in voltage profiles comes from the IBDGs Volt/VAR control, which considers the inverter's physical constraints. The other model merely injects the inverter's rated maximum power by default. 

\section{Conclusion and Future Work}
This work presented an improved model for an inverter-based distributed generator that accounts for current limits and implicit Volt/VAR control. Using a model that more closely approximates the inverter's control system can ensure that results are closer to the grid's true state. The approach's benefits are most evident when studying higher-penetration renewable systems where non-analytical models may overestimate or underestimate nodal voltages due to the absence of the natural incorporation of physical inverter limits. 

Today's models are insufficient under deterministic scenarios. They will most certainly underperform in future distribution grids, where the uncertainty from renewable generation is higher. Future work will consider how to represent uncertainty in renewable generation models and developing probabilistic analysis methods.

\bibliography{references}
\bibliographystyle{ieeetr}

\end{document}